\documentclass{article}
\usepackage{spconf,amsmath,graphicx}
\usepackage{amssymb}
\usepackage{bm}
\usepackage{booktabs}
\usepackage{csquotes}
\usepackage{dirtytalk}
\usepackage[hidelinks]{hyperref}
\usepackage{microtype}

\renewcommand{\paragraph}[1]{\noindent\textbf{#1}}

\title{SkinAugment: Auto-Encoding Speaker Conversions for Automatic Speech Translation}
\name{Arya D. McCarthy$^{\star \dagger}$ \qquad Liezl Puzon$^{\dagger}$ \qquad Juan Pino$^{\dagger}$}

\address{$^{\star}$ Center for Language and Speech Processing, Johns Hopkins University \\
    $^{\dagger}$ Facebook%
}

\begin{document}
\maketitle
\begin{abstract}
We propose autoencoding speaker conversion for training data augmentation in automatic speech translation. This technique directly transforms an audio sequence, resulting in audio synthesized to resemble another speaker's voice.
Our method compares favorably to SpecAugment on English--French and
English--Romanian automatic speech translation (AST) tasks as well
as on a low-resource English automatic speech recognition (ASR) task.
Further, 
in ablations, we show the benefits of both quantity and diversity in augmented data.
Finally, we show that we can combine our approach with augmentation by machine-translated transcripts to obtain a competitive end-to-end AST model  that outperforms a very strong cascade model on an English--French AST task.
Our method is sufficiently general that it can be applied to other speech generation and analysis tasks.

\end{abstract}
\section{Introduction}
\label{sec:intro}

The rarity of organic training examples presents a dilemma for automatic speech translation (AST); present-day AST is a low-resource task. While \textbf{end-to-end} models seem preferable from the perspective of inference latency or error propagation, they are difficult to train to competitive levels of performance. By contrast, \textbf{cascade} models \cite{post2013improved,6854197} are not bound to audio samples and their translations. They can leverage large-scale automatic speech recognition (ASR) and machine translation (MT) training datasets. Is it possible to create an AST model with high performance while keeping the benefits of end-to-end systems?

Data augmentation is a common solution for low resource scenarios and has been explored for both ASR and AST.
One of the most recent and successful data augmentation methods,  SpecAugment~\cite{Park2019}, modifies the spectrogram with time warping, frequency masking and time masking. AST methods to leverage ASR and MT data include pretraining~\cite{bansal-etal-2019-pre}, multitask learning~\cite{weiss2017sequence} and weakly supervised data augmentation~\cite{jia2019leveraging, pino2019harnessing}.

In this work, we generate additional audio samples without requiring transcripts, using a recent neural voice conversion technique, \say{text-to-speech skins}~\cite{polyak2019tts}. Operating on the raw wav audio, it isolates the essential from the contextual aspects of speech, transferring essential aspects into a new voice~\cite{moulines1995voice}. We apply this method to samples of AST and ASR training data to produce new variants, in a process we call \textsc{SkinAugment}.
We additionally investigate neural speaker normalization based on the same conversion model.

We assess our proposals on English--French and English--Romanian AST tasks as well as on a low-resource English ASR task.
We compare \textsc{SkinAugment} to
SpecAugment.

We find that \textsc{SkinAugment} effectively improves the performance of end-to-end AST models, without requiring additional annotated AST or MT data.
Particularly, we see BLEU gains of 2.2 on En--Fr and 3.3 on En--Ro.
\textsc{SkinAugment} outperforms both SpecAugment and a simple stochastic alteration we propose that improves SpecAugment on two AST tasks and one low-resource ASR task.
However, we find no significant benefit to using \textsc{SkinAugment} for test set normalization.

Further, we are able to produce a competitive end-to-end AST system by combining \textsc{SkinAugment} and weak supervision from machine-translated ASR samples.
This system outperforms a very competitive cascade~\cite{pino2019harnessing} by 1.1 BLEU.

\begin{figure}[t]
	\includegraphics[width=\linewidth]{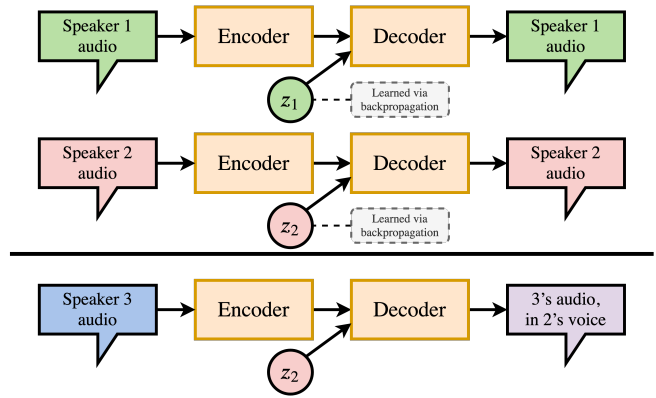}
	\caption{Conditional autoencoding of two speakers' audio (top). The latent speaker representation \(\boldsymbol{z}_i\) can transform (``skin'') new audio from unseen speakers (bottom).}
	\label{fig:autoencoding}
\end{figure}

\section{SKINAUGMENT: Augmentation with Voice Conversion by  Conditioned Autoencoding
}
Our speaker conversion technique~\cite{polyak2019tts} employs a convolutional wav-to-wav network, summarized in \autoref{fig:autoencoding}. The end-to-end encoder--decoder architecture optimizes an autoencoding loss, reproducing (a shifted version of) the original input while conditioned on a latent speaker representation. This representation is learned by backpropagation while optimizing the cross-entropy \(\ell\) over \(N\) training samples \(\boldsymbol{x}^{(i)}\):
\begin{equation}
\mathcal{L}(\theta) = \sum_{i=1}^{N}
	\ell\!\left(
		\mathrm{dec}\!\left(
			\mathrm{enc}(\boldsymbol{x}^{(i)}),
			\boldsymbol{f}_0(\boldsymbol{x}^{(i)}),
			\boldsymbol{z}_{s(i)}
		\right)\!,
		\boldsymbol{x}^{(i)}
	\right)\!
	\text{.}
\end{equation}
Here, \(s\) is a function that maps training indices to speaker IDs. \(\boldsymbol{Z} \in \mathbb{R}^{|S| \times d}\) is a matrix with \(d\)-dimensional latent representation for each of the \(|S|\) speakers seen during training. 
Extracting the fundamental frequency series with \(\boldsymbol{f}_0\) helps to preserve the original audio's prosody. We convert to new speakers by priming the decoder with the intended speaker's embedding. Training does not require parallel audio recordings between speakers, nor does it require transcripts of the audio. New speakers can easily be introduced by fine-tuning the model, conditioned on a new representation.

The method achieves competitive performance on the Voice Conversion Challenge 2018 benchmark \cite{Lorenzo-Trueba2018}, despite using fewer parameters than winning systems. While the method's value to the voice conversion task has been demonstrated \cite{polyak2019tts}, we show its utility for achieving superior %
\emph{downstream} performance. (A related spectrogram-to-spectrogram voice converter has been applied to speech separation, trained on approximately 150 times as many hours of data \cite{biadsy2019parrotron}. In principle, this method or others could also be employed as the voice conversion subcomponent in \textsc{SkinAugment}.)

\paragraph{Augmentation Policy.} One may ask whether sheer quantity of data or its diversity contributes more to performance: Is it helpful to hear diverse variants of the same audio? 
Our augmentation procedure, \textsc{SkinAugment}, lets us address this: We sample a fraction of the training data, then skin this subset into any of \(K\) arbitrarily chosen voices. In this work, we experiment with up to 16 skinned variants of the training data, sampling between 10\% and 100\% of the data to be skinned.

\section{Experimental Setup}

\subsection{Datasets and Evaluation}
An AST dataset pairs source-language audio with a target-language translation. 
We experiment on two standard AST datasets: AST LibriSpeech~\cite{kocabiyikoglu-etal-2018-augmenting} (English--French; we use the same setup as \cite{berard2018end}) and MuST-C (English--Ro\-man\-ian; 432 hours) \cite{mustc19}.
We also use AST LibriSpeech for low resource ASR. (Note that the AST LibriSpeech test sets do not correspond to the original LibriSpeech test sets.)
In all cases, we use the 200+ voices in LibriSpeech to train the conversion model; the LibriSpeech tasks let us evaluate conversions on \emph{in-domain} audio, while the MuST-C task evaluates conversions on \emph{out-of-domain} audio. As the AST LibriSpeech corpus's test set is a subset of LibriSpeech's training set, we remove all the AST LibriSpeech test set voices from LibriSpeech's training set before training the converter.

In later experiments, we further augment the training data by translating LibriSpeech's transcripts (removing test set occurrences~\cite{pino2019harnessing}) with an MT system. The MT system is trained on two standard datasets: WMT16 for En--Ro (600k sentence pairs) and WMT14 for En--Fr (29 million sentence pairs).

Our En--Fr AST cascade baseline's MT subsystem is trained on the same WMT corpora. The ASR subsystem is trained on the full LibriSpeech corpus. %

For AST, we report BLEU~\cite{papineni-etal-2002-bleu}
on tokenized output. (On the ASR task, the transcript is already tokenized; on the AST tasks, we tokenize translations with Moses~\cite{koehn-etal-2007-moses}.) For ASR, we use word error rate (WER), also on tokenized output.

\subsection{Model Architecture}
\label{sec:model_arch}

All of our experiments use the same mixed convolutional-recurrent end-to-end model architecture for conditional sequence generation, our focus being
data augmentation techniques. (Recent work suggests that AST performance with Transformer is similar to AST performance with this style of model  \cite{karita2019comparative}.)
We use a speech encoder consisting of two non-linear layers followed by two convolutional  layers  and  three  bidirectional  LSTM  layers,  along with  a custom  LSTM  decoder~\cite{berard2018end, pino2019harnessing}.
The encoder uses 40 log-scaled mel spectrogram features.
We use 3 decoder layers as in \cite{pino2019harnessing}, who report the number of parameters in each model. 

\textsc{SkinAugment} couples an off-the-shelf, fixed time-delay neural network (TDNN) encoder with a learned Wave\-Net decoder. Hyperparameters are as in \cite{polyak2019tts}.

\subsection{Baselines}
\paragraph{Cascade.}
We compare our data-augmented end-to-end model to a baseline cascade model. The ASR model is described in \autoref{sec:model_arch}, while the MT model uses a Transformer, trained on the WMT14 En--Fr parallel data. %
It achieves top performance on the AST LibriSpeech dataset of 21.3 BLEU~\cite{pino2019harnessing}. We also compare to \cite{berard2018end}'s cascade which lacks additional data.
 
\paragraph{SpecAugment.}
We also compare our end-to-end model with another popular data augmentation strategy, spectral augmentation (SpecAugment). SpecAugment adds perturbations at the feature level, whereas \textsc{SkinAugment} operates at the raw wave level.  We use the \emph{LibriSpeech double} setting \cite{Park2019}. 

\paragraph{SpecAugment-\(\bm{p}\).}
Further, we introduce a simple but effective variant: SpecAugment-\(p\), which applies SpecAugment to each batch with probability \(p\). (The standard SpecAugment would thus use \(p=1\).)
We found that SpecAugment with \(p=0.5\)
was effective in our setup.

\subsection{Augmentation and Normalization Settings}

We perform conditional generation of new data with either 8 or 16 voice conversions, applying them to 10\%, 25\%, 50\%, and 100\% of the training corpus. This creates transformed variants of our dataset in distinct (arbitrarily selected) voices. Generation is performed offline---thus not a prerequisite for inference---and is agnostic toward the AST model. While future work can explore a greater number of voices, we found this prohibitive in terms of training time. 

We compare these settings to standard SpecAugment, as well as to SpecAugment-\(p\) with \(p=0.5\), which we found to be effective.

Perhaps rather than making the AST model robust to the niceties of individual speakers' voices, we ought to eliminate those niceties. 
To test the effectiveness of translation on a consistent voice, we convert the test set to entirely be of the voice of one speaker and evaluate the BLEU score separately on these single-speaker skinned test sets. We select 8 voices arbitrarily. We then produce 8 such skinned test sets with \textsc{SkinAugment}, reporting average performance and standard deviation across the variants.

\subsection{Machine-Translated Augmentation} Existing AST samples are rare, leading research to explore avenues for weak supervision. Among these, machine-translated transcripts of large ASR corpora dramatically increase the performance of AST models \cite{jia2019leveraging,pino2019harnessing}. We therefore translate LibriSpeech transcripts with our Transformer, then concatenate these synthetic training instances to the AST data. We apply 16 skins to 25\% of the AST training data, as we found this to perform best.

\subsection{Training Settings}

We use the Adam optimizer~\cite{kingma2014adam} with a learning rate of 0.001 and gradient clipping of 5. The minibatch size is 96,000 frames. 
All experiments are conducted on 8 \textsc{Nvidia} Tesla V100 GPUs.
In order to compensate for the imbalance between
synthetic skinned data and original data, all models are fine-tuned on the original data for 40 epochs after convergence on the augmented data. We found a consistent improvement from fine-tuning.

We decode with a beam size of 20. To balance between the data sparsity of a word-level model and the training time of a character-level model, we use a SentencePiece \cite{kudo2018sentencepiece} unigram model with vocabulary size 10,000.

\section{Experimental Results}

\begin{figure}
    \centering
    \includegraphics[width=0.9465\linewidth]{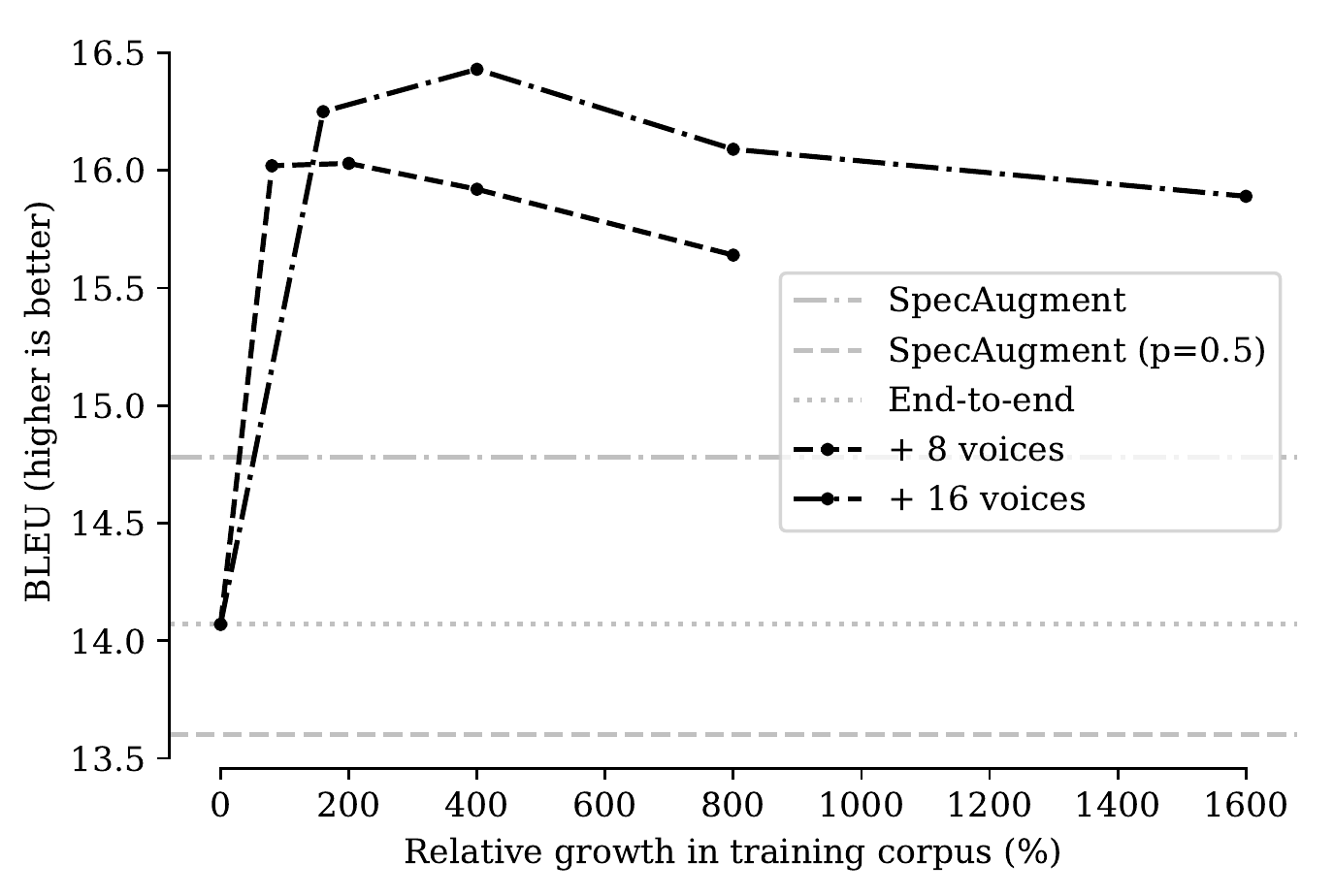}
    \caption{English--Romanian AST with out-of-domain skins. \textsc{SkinAugment} outperforms both SpecAugment variants.}
    \label{fig:en-ro}
\end{figure}

\begin{figure}
    \centering
    \includegraphics[width=0.9465\linewidth]{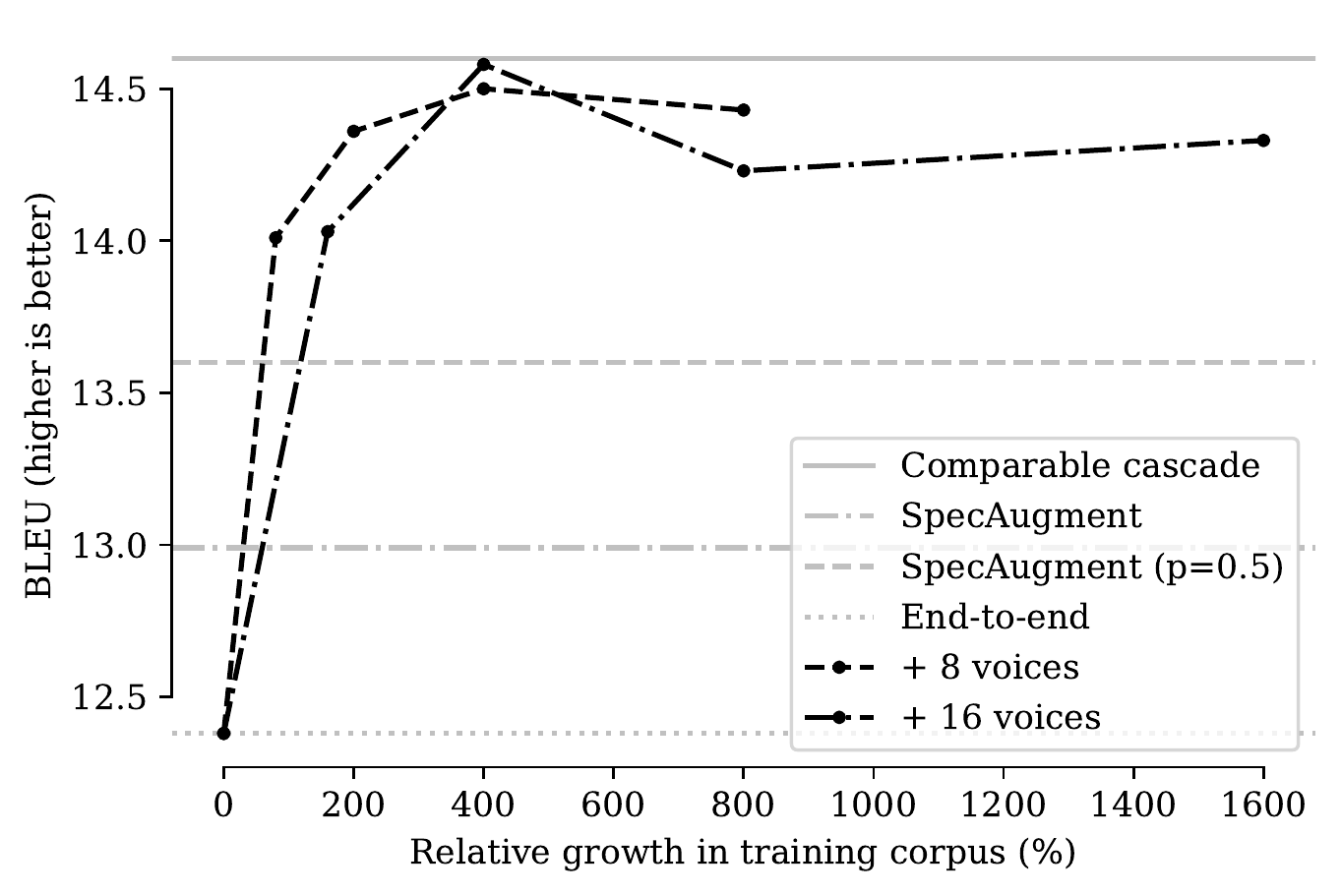}
    \caption{English--French AST with in-domain skins. Additional synthetic data is eventually harmful.}
    \label{fig:en-fr}
\end{figure}

\begin{figure}
    \centering
    \includegraphics[width=0.9465\linewidth]{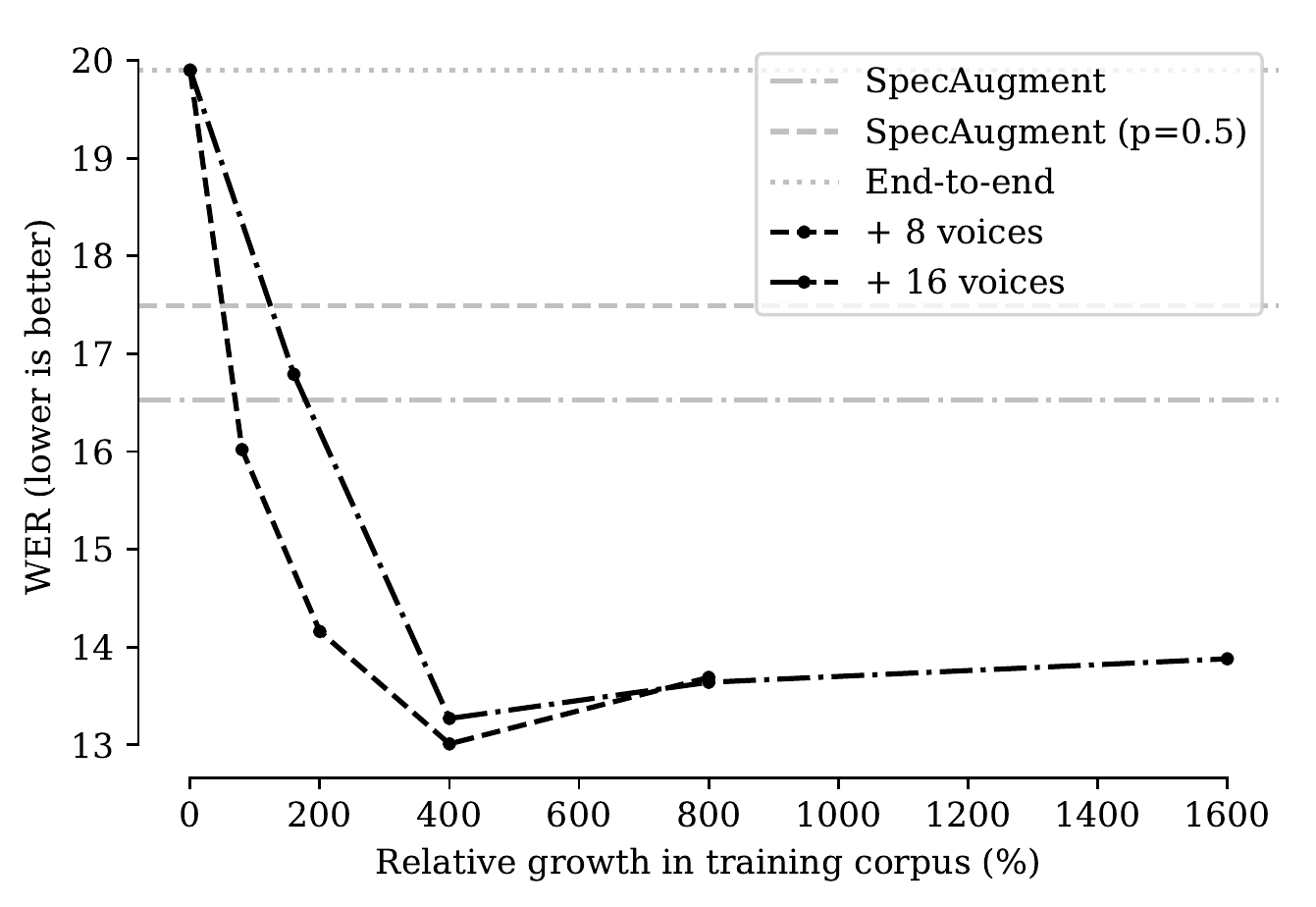}
    \caption{\textsc{SkinAugment} outperforms both variants of SpecAugment on the simulated low-resource ASR scenario.}
    \label{fig:asr}
\end{figure}

Results are presented in \autoref{fig:en-ro} (En--Ro AST), \autoref{fig:en-fr} (En--Fr AST), and \autoref{fig:asr} (ASR).
On all three tasks, we find that our augmentation strategy outperforms SpecAugment and SpecAugment-\(p\). We also found that SpecAugment-\(p\) outperforms SpecAugment on all tasks except ASR.

\textsc{SkinAugment} improves BLEU by  3.3 points for En--Ro and 2.2 for En--Fr over the end-to-end baseline. Our score of 14.58 matches the reported En--Fr score of \cite{berard2018end} with a cascade model (14.6), up to their reported significant figures.

\subsection{Quantity versus diversity of the augmented data} 
How much augmented data is needed for strong performance?
Does the advantage in the previous subsection come from pure quantity of data, or is the diversity of speakers advantageous?
We find that the performance of the end-to-end model tracks remarkably well with the \emph{amount} of data added, regardless of whether it comes from eight skins or sixteen, up to about twice the original size of the training data.

Beyond this, having more skins seems to be beneficial. When the amount of skinned data is 4\(\times\) the size of the training set, the 16-skin model (skinning 25\% of the training data) has a relative gain from the end-to-end baseline of 58\%, compared to the 8-skin model (skinning 50\% of the training data). Above this training set size (up to 16\(\times\)), performance begins to degrade. Nevertheless, it continues to outperform the baseline end-to-end model in our setting. For ASR and En--Ro AST, performance plateaus.

\subsection{Test set normalization}
For all three tasks, we also skin the test set to a single training-set voice, then evaluate. The motivation is to reduce variation in test data. To avoid reporting fortuitous but unrepresentative performance from a particular voice, we consider mean BLEU and standard deviation across 8 voices. Here, we report a negative result. In every case, the score on the unmodified test set is within one standard deviation of the voice-normalized mean. In \autoref{sec:mt-aug}, we find that translating normalized variants underperforms translating on the original audio by 0.6 BLEU on average, suggesting that the massive amount of unskinned audio obviates the benefit of skinned test data.

Our findings mesh well with \cite{keskin2019measuring}, who found that their Cycle-GAN voice converter was harmful for test set normalization and had negligible value for data augmentation; nevertheless, in their case and ours, increased amounts of skinned data led to better performance on normalized test sets. Future work can explore whether fine-tuning to the normalization voice improves performance.

\subsection{Machine-Translated Data for Augmentation} \label{sec:mt-aug}
Thus far, our synthetic data on the AST task has been generated by transforming original AST samples with either \textsc{SkinAugment} or SpecAugment(-\(p\)). However, adding translated data as weak supervision in our low-resource scenario improves performance significantly. \autoref{tab:my_label} shows an ablation: incorporating \textsc{SkinAugment}, translated transcripts (\say{+ MT}), or both. Furthermore, we demonstrate performance when using the original AST corpus from the augmented LibriSpeech release ~\cite{kocabiyikoglu-etal-2018-augmenting}, i.e.\ removing the  off-the-shelf automatic translations added in \cite{berard2018end}'s dataset (\say{\(-\) AT}). We speculate that the abysmal performance of the baseline AST LibriSpeech \(-\) AT is due to data scarcity, and that removing automatic translations for + MT helps because they are of lower quality.

\begin{table}[]
    \caption{Value of machine-translated transcripts combined with \textsc{SkinAugment} on AST LibriSpeech. We use 16 skins applied to 25\% of the corpus.}
    \vspace{0.125em}
    \centering
    \begin{tabular}{l r} 
    \toprule
    Data & BLEU \\
    \midrule
        AST LibriSpeech & 13.24 \\  %
        + \textsc{SkinAugment} & 15.22 \\ %
        + MT & 19.71 \\  %
        + MT + \textsc{SkinAugment} & 20.19 \\  %
    \midrule
        AST LibriSpeech \(-\) AT & 1.81 \\ %
        + MT & 21.78 \\  %
        + MT + \textsc{SkinAugment} & 22.44 \\  %
    \midrule
        Cascade (with ASR and MT data) & 21.31 \\
    \bottomrule
    \end{tabular}
    \label{tab:my_label}
\end{table}

\section{Conclusion}

We have evaluated speaker conversions using conditioned autoencoding for AST and ASR data augmentation.
A wav-to-wav CNN architecture learns latent speaker representations.
Swapping in a new speaker representation converts the voice in the audio. 
This yields more source audio for a given example.
The method is applicable to both data augmentation during training and speaker normalization for generation. 

While this method relies on additional audio data to train
the speaker conversion, it does not rely on transcribed text, which makes it appealing for scaling to different languages  and in low-resource scenarios where annotation can be costly.
\textsc{SkinAugment} compares favorably to SpecAugment, a popular data augmentation method that operates at the feature level.
We were also able to effectively combine speaker conversion data with MT-augmented ASR data.
Still, when instead applied to the test set at inference time as voice normalization, we observe no significant change in BLEU.

Creating AST data by text-to-speech (TTS) synthesis of parallel text corpora has shown mixed results; while \cite{jia2019leveraging} found that adding a TTS system's outputs improved performance, \cite{pino2019harnessing} were unable to find additional gains. The promise of \textsc{SkinAugment} to produce variants of a given audio without transcripts suggests that it could apply to such TTS data. Future work will explore the application of this augmentation approach to improving the effectiness of TTS data for AST.

\section{Acknowledgments}
We thank Hyunbin Park, Adam Polyak, Xiaohui Zhang, and Weiyi Zheng for assistance in generating voice-converted data.

\clearpage
\bibliographystyle{IEEEbib}
\bibliography{refs}

\end{document}